\documentclass[12pt]{article}
\usepackage{amstex}

\columnwidth\textwidth

\begin{document}

\newcommand{\be}{\begin{equation}}
\newcommand{\fe}{\end{equation}}
\newcommand{\beann}{\begin{eqnarray*}}
\newcommand{\feann}{\end{eqnarray*}}
\newcommand{\bea}{\begin{eqnarray}}
\newcommand{\fea}{\end{eqnarray}}
\newcommand{\nn}{\nonumber}
\newcommand{\ben}{\begin{enumerate}}
\newcommand{\fen}{\end{enumerate}}
\newtheorem{df}{Definition}
\newtheorem{thm}{Theorem}
\newtheorem{lem}{Lemma}
\newtheorem{prop}{Proposition}
\begin{titlepage}

\noindent
\hspace*{11cm} BUTP-2000/24 \\
\vspace*{1cm}
\begin{center}
{\LARGE Embedding variables \\[0.5cm]
in finite dimensional models} 

\vspace{2cm}

M.~Ambrus
and
P. H\'{a}j\'{\i}\v{c}ek \\
Institute for Theoretical Physics \\
University of Bern \\
Sidlerstrasse 5, CH-3012 Bern, Switzerland \\
\vspace*{2cm}

December 2000 \\ \vspace*{1cm}

\nopagebreak[4]

\begin{abstract}
  Global problems associated with the transformation from the Arnowitt, Deser
  and Misner (ADM) to the Kucha\v{r} variables are studied. Two models are
  considered: The Friedmann cosmology with scalar matter and the torus sector
  of the 2+1 gravity. For the Friedmann model, the transformations to the
  Kucha\v{r} description corresponding to three different popular time
  coordinates are shown to exist on the whole ADM phase space, which becomes a
  proper subset of the Kucha\v{r} phase spaces.  The 2+1 gravity model is
  shown to admit a description by embedding variables everywhere, even at the
  points with additional symmetry. The transformation from the Kucha\v{r} to
  the ADM description is, however, many-to-one there, and so the two
  descriptions are inequivalent for this model, too. The most interesting
  result is that the new constraint surface is free from the conical
  singularity and the new dynamical equations are linearization stable.
  However, some residual pathology persists in the Kucha\v{r} description.
\end{abstract}

\end{center}

\end{titlepage}

\section{Introduction}
\label{sec:intro}
Generally covariant systems are quite popular in the theoretical physics of
today. Each such model contains one or more spacetime-like objects. For
example, in string theory, we find the target spacetime as well as string (and
membrane) sheets. The variables that specify points in the phase space are
then tensor (density) fields on Cauchy surfaces in some of the spacetimes. For
example, in general relativity, the first and second fundamental forms of the
Cauchy surface are used, or rather some modifications thereof, the so-called
ADM variables $q_{kl}(x)$ and $\pi^{kl}(x)$ \cite{ADM}. We call the canonical
formalism based on these variables {\em ADM description}.

As early as 1962, it was recognized \cite{bayer} that the ADM variables
contain a mixture of two types of information. The first has to do with the
physical, gauge independent state of the system. The second just tell us where
in the spacetime does the Cauchy surface lie.

The mathematical language of this idea has been worked out by Kucha\v{r}
\cite{K1}. The variables that describe the position of the Cauchy surface are
the so-called {\em embeddings}: maps of the form $X : \Sigma \mapsto {\mathcal
  M}$ of the Cauchy manifold $\Sigma$ into the spacetime manifold $\mathcal
M$. The gauge invariant, true physical degrees of freedom can be described by
variables of the so-called Heisenberg picture \cite{K2}. They are observables
in the sense of Dirac \cite{D}. The momenta $P$ conjugate to the embeddings
$X$ are simultaneously the new constraint functions. We call the canonical
formalism based on these variables {\em Kucha\v{r} description}.

One advantage of Kucha\v{r} variables is that they enable a
four-dimensional, spacetime, formulation of the canonical theory: all Cauchy
surfaces are admitted in the canonical description of the dynamics (``bubble
time'', or ``many-finger time'', \cite{K1}). This is to be compared with the
one-parameter time evolution based on a particular choice of a one-dimensional
family of Cauchy surfaces in each solution spacetime, the so-called foliation.
A foliation is a particular 3+1 split of the four-dimensional spacetime. The
original ADM reduction program (for a today's version, cf.\ \cite{FM}) was
based on such a split. Kucha\v{r}' approach allows to write down explicitly
the action of the four-dimensional diffeomorphism group \cite{IK}---the gauge
group of the model.

A canonical transformation from the ADM to the embedding variables, their
conjugate momenta and the observables will be called {\em Kucha\v{r}
  decomposition} or {\em Kucha\v{r} transformation}. The Kucha\v{r}
transformation turned out to be a difficult task. It was managed only for few
special models (\cite{K3}, \cite{K4}, \cite{BK} and \cite{KRW}). Moreover,
some general, negative results were published. In \cite{H1} and \cite{H2},
simple models were constructed that did not allow a global Kucha\v{r}
decomposition. Torre \cite{T1} showed that the decomposition, which, in fact,
brought the system to the form of the so-called ``already parameterized
system'', was impossible at some points of the constraint surface of general
relativity. These were the points that, as Cauchy data, evolved to spacetimes
with additional Killing vectors. Thus, even the existence of Kucha\v{r}
decomposition was questioned.

Some progress in this situation has been achieved in \cite{HKi} (see also
\cite{Hkief}). The conditions for the existence of the Kucha\v{r}
transformation have been clarified. First, each Kucha\v{r} decomposition is
associated, and in fact determined, by a choice of gauge. The Kucha\v{r}
coordinate chart can cover only such part of the constraint surface, for which
a common gauge fixing exists. Second, all points of the constraint surface
must be excluded that evolve to spacetimes with {\em any} isometries, not just
with Killing vectors. And, finally, even if these conditions are satisfied,
the existence could only be shown for a neighbourhood of the constraint
surface, not for the whole ADM phase space.

The aim of the present paper is to start studying the conditions mentioned in
the previous paragraph. This would be rather difficult in general context.  We
shall, therefore, start by studying finite-dimensional, ``minisuperspace''
models. For such models, the spacetime manifold $\mathcal M$ is effectively
one-dimensional and the Cauchy manifold is just a point, so the space of
embeddings can be identified with $\mathcal M$---a finite-dimensional space.
The models chosen are completely solvable. This enables us to construct
Kucha\v{r} transformations explicitly (the proof in \cite{HKi} is not
constructive).

The plan of the paper is as follows. In Sec.~\ref{sec:fried}, we consider the
Friedmann cosmological model driven by a zero-rest-mass, conformally coupled
scalar field. This model has been studied in some detail in \cite{claus}.
First, we specify the gauge needed for Kucha\v{r} transformation. In the
one-dimensional spacetime model, it can be called ``choice of time''; the time
coordinate is, in fact, an embedding variable. It is advantageous to decompose
the choice into two steps. The first one specifies the lapse $N$ as a function
on the ADM phase space $\mathcal P$. Then the canonical equations of the
Hamiltonian $P := N{\mathcal H}$ define the so-called {\em trajectories}
everywhere in $\mathcal P$; $\mathcal H$ is the constraint function. The
second step is a choice of the surface transversal to the trajectories as
the origin of time. We study three choices of time: {\em conformal}, {\em
  proper} and {\em CMC time} and try to find the corresponding Kucha\v{r}
coordinates on the whole of $\mathcal P$.

The model of Sec.~\ref{sec:torus} is the torus sector of the 2+1 gravity
theory, partially reduced so that a minisuperspace model results. This has
been carried out in \cite{M1}, from where we adopt our starting formulae. The
model is interesting for several reasons. Its constraint set $\mathcal C$ does
contain points associated with higher symmetry---the static tori.  $\mathcal
C$ has a bifurcation and conical singularity at these points. The conical
singularity is a feature associated with additional Killing vectors, see
\cite{FMM}. It is also the cause of the so-called {\em linearization
  instability} \cite{M3}. $\mathcal C$ has no well-defined differential
structure at these points. This is a difficulty not only for the
transformation to Kucha\v{r} variables, but also for the definition of the ADM
physical phase space (see \cite{M2}). Finally, this model does not admit a
globally transversal surface.  We can, therefore, study this topological
obstruction, too.

All these problems disappear if we truncate the model by excising the points
associated with the static tori as it has been done in \cite{M1} and
\cite{H2+1}. The truncated model consists of two separated parts. Each part
admits a globally transversal surface, a global chart of Kucha\v{r} variables,
and a nice physical phase space. In the present paper, we are trying to extend
both parts of the truncated model.

Sec.\ \ref{sec:physph} investigates important properties of the physical phase
space of the extended model. We construct an atlas for the physical phase
space from a chosen family of transversal surfaces in the constraint set.  In
this way, a smooth manifold (in fact, analytic) can be obtained.

In Sec.\ \ref{sec:ev}, we turn to the embedding variables. Strictly speaking,
the negative results of \cite{T1} and \cite{HKi} only imply that the ADM
variables cannot be transformed into Kucha\v{r} ones at the points with higher
symmetry.  This does not mean that there is no Kucha\v{r} description
including the solutions with additional symmetry. However, if it exists it
cannot be equivalent to the ADM description. 

Our atlas for the physical phase space serves as a starting point. The
transversal surfaces defining it can be extended from the constraint surface
to a part of the ADM phase space.  There is a patch of Kucha\v{r} coordinates
for each transversal surface. In this way, we obtain a Kucha\v{r} description
of the whole model. The transformation from the Kucha\v{r} description to the
ADM one becomes singular, many-to-one, at the points of higher symmetry (the
trajectories containing these points are zero-dimensional in ADM, and
one-dimensional in Kucha\v{r} description). However, all classical solutions
of the new system coincide completely with the corresponding solutions of the
old one. Yet the new constraint surface is free from the bifurcation and the
conical singularity. The new dynamical equations are {\em linear}. This has an
obvious but amusing consequence: they are linearization stable.

The results are discussed in Sec.\ \ref{sec:concl}. There is also an attempt
at a synthesis of the results from both minisuperspace models.

\section{Friedmann model \newline with conformally coupled scalar field}
\label{sec:fried}
In this section, we shall study the spatially closed Friedmann cosmological
model with a particular matter content: a rest-mass-zero, conformally coupled
scalar field. 

The action has the form
\[
  S = \int dt(p_a\dot{a} + p_\phi\dot{\phi} - N {\mathcal H}),
\]
where $a(t)$ is the scale factor of the Robertson-Walker line element,
\be
 ds^2 = - N^2 dt^2 + a(t)^2\left(\frac{dr^2}{1-r^2} +
   r^2(d\vartheta^2 + \sin\vartheta^2 d\varphi^2)\right),
\label{1;1}
\end{equation}
$N$ is the lapse function, $\phi$ is defined in terms of the original
scalar field $\Phi$ by
\[
  \phi := \frac{2\sqrt{2}\text{G}}{3}a\Phi,
\]
G is the Newton constant, and $\mathcal H$ is the Hamiltonian constraint,
\[
  {\mathcal H} = \frac{1}{2a}(-p_a^2 + p_\phi^2 - a^2 + \phi^2).
\]
For more details see \cite{claus}.

The ADM phase space $\mathcal P$ is four-dimensional, covered by the canonical
chart $a$, $\phi$, $p_a$ and $p_\phi$ with ranges
\begin{alignat*}{2}
  a & \in (0,\infty), & \qquad \phi & \in (-\infty,\infty), \\
  p_a & \in (-\infty,\infty), & \qquad p_\phi & \in (-\infty,\infty).
\end{alignat*}
The constraint surface $\mathcal C$ is the three-dimensional ``cone'',
\[
  -p_a^2 + p_\phi^2 - a^2 + \phi^2 = 0.
\]
The background manifold $\mathcal M$ is one-dimensional, ${\mathcal M} =
{\mathbf R}$. For its complete definition, a choice of time coordinate $T$ is
needed \cite{HKi}. The Cauchy manifold is represented by a zero-dimensional
manifold (a point) $\Sigma$, and the space Emb$(\Sigma,{\mathcal M})$ of
embeddings $T : \Sigma \mapsto {\mathcal M}$ can be identified with ${\mathcal
  M}$:
\[
  \text{Emb}(\Sigma,{\mathcal M}) = {\mathcal M}.
\]

A set of Kucha\v{r} variables consists then of the time variable $T$, its
conjugate momentum $P$, which is proportional to the Hamiltonian constraint,
and two Dirac observables, which are constants of motion. For the present
case, it is not difficult to find the transformation to such variables if $T$
is chosen so that the equations of motion simplify.

\subsection{Conformal time}
\label{sec:conf}
A suitable choice of time is connected with the following value of the lapse:
\be
  N = a.
\label{4;1}
\end{equation}
Equation (\ref{1;1}) shows that $T$ is a {\em conformal time} then. The
conjugate variable is
\begin{equation}
  P := N{\mathcal H} = \frac{1}{2}(-p_a^2 + p_\phi^2 - a^2 + \phi^2).
\label{P}
\end{equation}
The time coordinate is not yet completely specified. Some surface is to be
chosen as the origin $T=0$. 

The equations of motion corresponding to the Hamiltonian $P$ are:
\begin{alignat}{2}
  \dot{a} & = -p_a, & \qquad \dot{p}_a & = a, 
\label{xi1} \\
  \dot{\phi} & = p_\phi, & \qquad \dot{p}_\phi & = -\phi.
\label{xi2}
\end{alignat}
It follows that $\dot{p}_a$ is positive everywhere in $\mathcal P$, and we can
choose the surface defined by $p_a = 0$ as $T = 0$. The resulting general
solution to the equations of motion is:
\begin{alignat}{2}
  a & = A\cos T, & \qquad  p_a & = A\sin T, 
\label{5;1} \\
  \phi & = B\cos(T+C), & \qquad p_\phi & = -B\sin(T+C),
\label{5;2}
\end{alignat}
where $A$, $B$ and $C$ are constants. We can express $P$ by these constants:
\[
  P = \frac{1}{2}(B^2 - A^2).
\]
The functions $T$, $P$, $B$ and $C$ form a complete set of independent
variables. Eqs.~(\ref{5;1}) and (\ref{5;2}) can be written by means of these
variables if $A = \sqrt{B^2 - 2P}$ is substituted for $A$. They can then be
considered as transformation equations from the variables $a$, $p_a$, $\phi$
and $p_\phi$ to $T$, $P$, $B$ and $C$. Let us express the Liouville form in
terms of the new variables. A simple calculation reveals:
\beann
  \lefteqn{p_a da + p_\phi d\phi = P dT + \frac{1}{2}B^2 dC}  \\
  && + d\left(-P\sin T\cos T + \frac{1}{2}B^2\sin T\cos T -
  \frac{1}{2}B^2\sin(T+C)\cos(T+C)\right). 
\end{eqnarray*}
To improve the right-hand side, we introduce the functions $q$ and $p$ by
\[
  q = B\cos C, \quad p = -B\sin C;
\]
this implies that
\[
  pdq = \frac{1}{2}B^2 dC + d\left(-\frac{1}{2}B^2\sin C\cos C\right).
\]
Hence,
\begin{alignat}{2}
  a & = \sqrt{q^2+p^2-2P}\cos T, & \qquad p_a & = \sqrt{q^2+p^2-2P}\sin T, 
\label{7;1} \\
  \phi & = q\cos T + p\sin T, & \qquad p_\phi & = -q\sin T + p\cos T,
\label{7;2}
\end{alignat}
is a canonical transformation. The meaning of the variables $q$ and $p$ can be
inferred from Eq.\ (\ref{7;2}): $q = \phi\vert_{T=0}$ and $p =
\pi_\phi\vert_{T=0}$. These are values of the field $\phi$ and its momentum
$\pi_\phi$ at the surface of maximal expansion.

The transformation defined by Eqs.\ (\ref{7;1}) and (\ref{7;2}) maps the
following subset of ${\mathbf R}^4$ with natural coordinates $T$, $P$, $q$ and
$p$ onto $\mathcal P$:
\begin{equation}
  (q,p) \in {\mathbf R}^2\setminus \{0\},\quad T \in
  \left(-\frac{\pi}{2},\frac{\pi}{2}\right),\quad P \in
  \left(-\infty,\frac{1}{2}(q^2+p^2)\right).
\label{7;3} 
\end{equation}
$T$ is the embedding variable corresponding to our choice of gauge. Its
conjugate variable $P$ is proportional to the constraint function. The
remaining variables $q$ and $p$ are Dirac observables. They span the physical
phase space $\Gamma$. Hence, the new action reads
\begin{equation}
  S = \int dt\left(P\dot{T} + p\dot{q} - N'P\right),
\label{kuch}
\end{equation}
where $N' = aN$ is the new lapse function. The action has the
Kucha\v{r} form.

The boundaries defined by Eq.~(\ref{7;3}) have the following meaning. $T =
-\pi/2$ is the big bang and $T = \pi/2$ the big crunch singularity of the
solution to Einstein equations for our model, if $P = 0$. The points are still
singular for $P \neq 0$, but this is a property of the present gauge ($N$ can
be chosen such that the solutions of the resulting equations of motion outside
the constraint surface are regular). The boundary $P = (q^2 + p^2)/2$
corresponds to $a(T) = 0$ for all $T$. This ``solution'' does not define any
spacetime. Finally, the point $q = p = 0$ corresponds to the scalar field
being identically zero. Then, again, there is no spacetime solution for $P =
0$.

The existence of bounds on the embeddings and their conjugate momenta seems to
be an important general feature of Kucha\v{r} transformation. Ref.~\cite{HKi}
already mentioned one kind of such bound: the embeddings must be everywhere
space-like for each given geometry. In the present case, only very special
embeddings are allowed, which are automatically space-like. On the other hand,
our findings on the bound that must be satisfied by $P$ is rather unexpected
and new. To understand it, let us recall that a Kucha\v{r} transformation is
described in \cite{HKi} as a map $\chi : \Gamma \times T^*\text{Emb}(\Sigma,
{\mathcal M}) \mapsto {\mathcal P}$. $\chi$ is a symplectic diffeomorphism and
its existence has been shown (under certain conditions) only in an open subset
$\mathcal U$ of $T^*\text{Emb}(\Sigma,{\mathcal M})$ such that $\chi(\mathcal
U)$ is a neighbourhood of the constraint surface $\mathcal C$ in $\mathcal P$.
One would expect that $\chi(\mathcal U)$ is a proper subset of $\mathcal P$ so
that the transformation exists only for limited values of the ADM variables,
because nothing more has been proved in \cite{HKi} but there is still some
uncertainty. On the other hand, $\mathcal U$ must be a proper subset of
$T^*\text{Emb}(\Sigma,{\mathcal M})$, so there are always some bounds on $X
\in \text{Emb}(\Sigma,{\mathcal M})$ and $P \in
T^*_X\text{Emb}(\Sigma,{\mathcal M})$.

In our case, Emb$(\Sigma,{\mathcal M}) = {\mathcal M} = {\mathbf R}$, and we
also use the letter $T$ rather that $X$ to denote an embedding. Then
$T^*\text{Emb}(\Sigma,{\mathcal M}) = {\mathbf R}$. Our result is that
$\chi(\mathcal U) = {\mathcal P}$ so that there are only bounds on $P$ and
$T$, not on the ADM variables. The interpretation is that the whole ADM phase
space $\mathcal P$ is a proper subspace of the Kucha\v{r} phase space $\Gamma
\times T^*\text{Emb}(\Sigma,{\mathcal M})$.

Let us observe that the points of $\Gamma \times
T^*\text{Emb}(\Sigma,{\mathcal M})$ that do not satisfy the bound (\ref{7;3})
for $P$ do not define any reasonable initial data for the spacetime and the
scalar field. However, one can use the action (\ref{kuch}) in the whole space
$\Gamma \times T^*\text{Emb}(\Sigma,{\mathcal M})$ without any harm. All
points of the constraint surface satisfy the bounds, so the solution of the
equations of motion within the ADM framework coincide with those within the so
extended Kucha\v{r} framework.

\subsection{Transversal surface}
\label{sec:transvers}
Let us study the geometrical structures that underlie the calculation of 
the previous section. 

The first step has been a choice of fixed phase-space function for the lapse
$N$. This has determined the true Hamiltonian $P$ by Eq.\ (\ref{P}). The
Hamiltonian vector field $\xi_P$ is given by the right-hand sides of Eqs.\ 
(\ref{xi1}) and (\ref{xi2}). It is important to observe that the direction of
$\xi_P$ is independent of $N$ at the constraint surface $\mathcal C$; it
is only the parametrization of the integral curves of $\xi_P$ that changes
with $N$. Outside $\mathcal C$, however, even the direction of $\xi_P$
depends on $N$, and the resulting integral curves form different
foliations of $\mathcal P$ for different $N$. 

The variable $T$ is to be conjugate to $P$. This implies the condition
\begin{equation}
  \xi_PT = 1.
\label{11;1}
\end{equation}
Hence, any parameter of the integral curves of $\xi_P$ can be chosen as $T$.

Let us denote the remaining two variables that we are looking for by $X$ and
$Y$. They are to form a canonical chart together with $T$ and $P$. It follows
that they have vanishing Poisson brackets with $P$:
\begin{equation}
  \xi_PX = \xi_PY = 0.
\label{11;2}
\end{equation}
Thus, $X$ and $Y$ are ``integrals of motion''. Observe that condition
(\ref{11;2}) depends on the choice of $N$ outside $\mathcal C$. At
$\mathcal C$, it is, however, independent of it, and it implies that $X$ and
$Y$ are Dirac observables. As also $\xi_PP = 0$, we conclude that the
functions $P$, $X$, and $Y$ form a complete set of independent integrals of
motion.

The conditions (\ref{11;1}) and (\ref{11;2}) do not determine the functions
$T$, $X$ and $Y$. We can fix $T$ using the following idea. Let $T$ be some
function satisfying Eq.\ (\ref{11;1}). Then the equation $T =$ const
defines a surface in $\mathcal P$ at least for some value of the constant.
This surface must intersect each integral curve of $\xi_P$ at most once (there
can be curves along which $T$ does not attain the value of the constant).
Moreover, the tangent spaces to the surface $T =$ const and that to the
integral curves of $\xi_P$ must have only the zero vector in common at every
point of the surface. We call a surface that satisfies both conditions {\em
  transversal}, and {\em globally transversal} if it intersects all integral
curves of $\xi_P$.  Suppose that the vector field $\xi_P$ admits a globally
transversal surface $\mathcal T$. Then the function $T$ can be chosen so that
it vanishes at $\mathcal T$; by that, the function is completely determined.

Let us turn to the functions $X$ and $Y$. They must have vanishing Poisson
brackets with $T$. Hence they have to satisfy the conditions
\begin{equation}
  \xi_TX = \xi_TY = 0.
\label{13;1}
\end{equation}
where $\xi_T$ is the Hamiltonian vector field of $T$. Observe that the Lie
bracket between $\xi_P$ and $\xi_T$ vanishes,
\[
  [\xi_T,\xi_P] = 0,
\]
because $\{T,P\} = 1$. Our construction of the functions $X$ and $Y$ is
based on this observation.

Let $\mathcal T$ be a globally transversal surface. Consider the
two-dimensional surface ${\mathcal T} \cap {\mathcal C}$. The pull-back
$\omega$ of the symplectic form $\Omega$ from $\mathcal P$ to ${\mathcal T}
\cap {\mathcal C}$ is again symplectic (non-degenerate). The symplectic
manifold $(\mathcal T \cap \mathcal C ,\omega)$ can be identified with the
physical phase space $\Gamma$.

Let us choose two coordinates $x$ and $y$ on ${\mathcal T} \cap {\mathcal C}$
satisfying
\[
  \{x,y\}_\omega = 1.
\]
We extend these functions in two steps to the whole of $\mathcal P$. First, we
use the condition (\ref{13;1}) to extend them to $\mathcal T$. Eq.\ 
(\ref{13;1}) can be considered as a differential equation on $\mathcal T$: as
$\xi_TT = 0$, the vector field $\xi_T$ is tangential to $\mathcal T$. Let the
functions $X$ and $Y$ at $\mathcal T$ satisfy the differential equations
(\ref{13;1}) together with the initial conditions
\[
  X\vert_{{\mathcal T}\cap{\mathcal C}} = x,\quad Y\vert_{{\mathcal
  T}\cap{\mathcal C}} = y. 
\]
This is sensible because the surface ${\mathcal T}\cap{\mathcal C}$ is
transversal to $\xi_T$ in $\mathcal T$. The reason is that $\xi_TP = -1$ and
${\mathcal T}\cap{\mathcal C}$ is defined by $P = 0$.

The second step is to use the differential equations (\ref{11;2}) with the
initial conditions at $\mathcal T$ given by the values of $X\vert_{\mathcal
  T}$ and $Y\vert_{\mathcal T}$ as obtained in the previous step. The two
steps result in a pair of functions $X$ and $Y$ that automatically satisfy the
remaining Poisson bracket conditions. This follows from the Jacobi identity as
follows.

At $\mathcal T$, we have
\begin{equation}
\begin{split}
  \xi_T\{X,Y\} & = \{\{X,Y\},T\} \\
  & = - \{\{T,X\},Y\} - \{\{Y,T\},X\} 
  = \{\xi_TX,Y\} - \{\xi_TY,X\} \\ & = 0
\end{split} 
\end{equation}
because of Eqs.\ (\ref{13;1}). Eq.\ (\ref{13;1}) implies that 
\begin{eqnarray}
  \{X,T\}\vert_{\mathcal T} & = & 0,
\label{XT} \\
  \{Y,T\}\vert_{\mathcal T} & = & 0,
\label{YT}
\end{eqnarray}
and $\{X,Y\}\vert_{\mathcal T}$ is constant along the integral
curves of $\xi_T$.

Similarly, $\xi_P\{X,Y\} = 0$, $\xi_P\{X,T\} = 0$, and $\xi_P\{T,Y\} = 0$ for
the propagation along $\xi_P$ in the second step. From the values given by
Eqs. (\ref{XT}) and (\ref{YT}), and from Eq.\ (\ref{11;2}), we obtain that
\begin{equation}
  \{X,T\} = \{Y,T\} = \{X,P\} = \{Y,P\} = 0
\label{st}
\end{equation}
everywhere in $\mathcal P$. We also have that $\{T,P\} = 1$, so it remains
only to show that $\{X,Y\} = 1$ everywhere. Now, $T$, $P$, $X$, and $Y$ are
independent functions in a neighbourhood of the surface ${\mathcal
  T}\cap{\mathcal C}$ and can be chosen as coordinates there. The only
non-zero components of the symplectic form $\Omega$ in these coordinates are
$\Omega_{PT} = -\Omega_{TP} = 1$ and $\Omega_{YX} = -\Omega_{XY}$ because of
Eq.\ (\ref{st}). The surface ${\mathcal T}\cap{\mathcal C}$ is defined by the
embedding relations
\[
  X = x,\quad Y = y,\quad P = 0,\quad T = 0.
\]
Hence, the pull-back $\omega$ is 
\[
  \omega_{yx} = \Omega_{YX}\vert_{P=T=0}
\]
and so $\{X,Y\}\vert_{{\mathcal T}\cap{\mathcal C}} = 1$ because $\omega_{yx}
= 1$. The desired result follows, for the bracket $\{X,Y\}$ must be constant
along $\xi_T$ and $\xi_P$.

Let us summarize. The construction of the previous section is based on three
choices: the function $P$, the transversal surface $\mathcal T$, and the
coordinates $x$ and $y$ on ${\mathcal T}\cap{\mathcal C}$. The Poisson
bracket conditions on the functions $T$, $P$, $X$, and $Y$ imply differential
equations that propagate the functions from ${\mathcal T}\cap{\mathcal C}$ to
$\mathcal T$ and from $\mathcal T$ to $\mathcal P$. The result is unique,
given the three choices.

The propagation from ${\mathcal T}\cap{\mathcal C}$ to $\mathcal T$ has been
only implicit in the calculation of Sec.\ \ref{sec:transvers} because the
functions $B$ and $C$ that satisfy Eqs.\ (\ref{13;1}) have been guessed.

From the commutativity of the vector fields $\xi_P$ and $\xi_T$, it follows
that the propagation is independent of the way chosen. Hence, we could
propagate first from ${\mathcal T}\cap{\mathcal C}$ to $\mathcal C$ along
$\xi_P$ and then from $\mathcal C$ to $\mathcal P$ along $\xi_T$ with the same
result. To do that, there has first to be given the function $T$ everywhere on
$\mathcal P$ instead of $P$. We can, therefore, call the method of the
previous section {\em P-way} and the alternative method {\em T-way}. We shall
use the $T$-way in the next section.

\subsection{Bergmann-Komar transformation}
\label{sec:B-K}
The conformal time $T$ of Sec.\ \ref{sec:conf} has a well-defined value at
each point of any spacetime solution. In general, the transformation from $T$
to some different ``time coordinate'' $T'$ will be solution dependent. For our
model, it can depend on the variables $P$, $q$ and $p$. Such transformations
have been introduced and studied by Bergmann and Komar \cite{B-K}.

As we have shown in Sec.\ \ref{sec:conf}, a choice of time can be done in two
steps: that of lapse function and that of transversal surface. If $N$ is not
changed, the Hamiltonian $P$ will be preserved. The solution arcs (the point
sets defined by the trajectories) will then be the same {\em everywhere} in
$\mathcal P$. A change of transversal surface leads only to a
reparametrization of the trajectories. Hence, such a transformation can be
considered as a genuine, solution-dependent, reparametrization of the
trajectories everywhere. In our case, it has the general form
\begin{alignat}{2}
  T' & =  T + \tilde{T}(P,q,p), & \qquad P' & = \tilde{P}(P,q,p), 
\label{20;1} \\
  q' & = \tilde{q}(P,q,p) , & \qquad p' & = \tilde{p}(P,q,p).
\label{20;2}
\end{alignat}
Such transformations form a subgroup. They do not change the character of the
time. In our case, it remains conformal time. Still, it is a Bergmann-Komar
transformation that cannot be implemented, in general, by a unitary
transformation in the quantum theory (see \cite{paris}).

If we change $N$, the character of time changes. In this section, we are
going to study two such examples: the {\em proper} and the {\em
  constant-mean-external-curvature} times. These are two relatively popular
choices in cosmology.

In fact, a change of $N$ leads to a more radical change of the trajectories
outside the constraint surface than just a reparametrization. The notion of
``solution spacetime'' is gauge independent only at $\mathcal C$.  Yet the
trajectories are important for us everywhere in $\mathcal P$: they define the
Kucha\v{r} coordinates there.

The trajectories are solutions to the canonical equations of the Hamiltonian
$P= N{\mathcal H}$. They are so uniquely defined through all points of
$\mathcal P$. The fact that the solution arcs depend on $N$ outside $\mathcal
C$ has to do with the way that dynamics of a generally covariant system is
usually formulated. The classical dynamics of such a system is completely
determined by the constraint surface (but cf.\ \cite{H-Kuch}). The form of the
constraint functions is irrelevant as long as they define the same constraint
surface.

How are the trajectories outside the constraint surface to be interpreted? To
be sure, each gauge and any of the corresponding trajectories define a unique
Robertson-Walker spacetime and scalar field on it. The gauge supplies the
lapse function $N(q,p,P,T)$ and then each of the corresponding trajectories
determines a unique scale factor $a(q,p,P,T)$ as well as the scalar field
$\phi(q,p,P,T)$. In this manner, there is a spacetime with the metric
\begin{equation}
  ds^2 = -N^2(q,p,P,T)dT^2 + a^2(q,p,P,T)\left(\frac{dr^2}{1-r^2} +
  r^2d\Omega^2\right)
\label{trajst}
\end{equation}
for each trajectory determined by the constant values of $q$, $p$ and $P$. The
scalar field $\phi(q,p,P,T)$ can be considered as a field on this spacetime.
Of course, the relation between the momenta $p_a$ and $p_\phi$ on one hand and
the velocities $da/dT$ and $d\phi/dT$ on the other will {\em not}, in general,
coincide with those obtained from the second order formalism, eg., by
\[
  p_a = \frac{\partial {\mathcal L}}{\partial \dot{a}}.
\]
This does not seem, however, to disturb the interpretation based on the
existence of the solution spacetime (\ref{trajst}).

\subsubsection{Proper time}
\label{sec:proper}
Here, we calculate the transformation from $T$, $P$, $q$ and $p$ to Kucha\v{r}
variables corresponding to the proper time $T'$.

The lapse function $N$ that is associated with the proper time has the
value $N = 1$. We obtain that $adT = dT'$. At the constraint surface,  $P
= 0$, and Eq.\ (\ref{7;1}) implies
\begin{equation}
  dT' = \sqrt{q^2 + p^2}\cos T dT.
\label{23;1}
\end{equation}
The $T'$-curves at the constraint surface consist of the same points as the
$T$-curves. Hence, the values of $q'$ and $p'$ are again constant along them,
$q' = q$ and $p' = p$ and Eq.\ (\ref{23;1}) has the integral
\begin{equation}
  T' = \sqrt{q^2 + p^2}\sin T,
\label{24;1}
\end{equation}
where we have chosen the same transversal surface ${\mathcal T}\cap{\mathcal
  C}$ as in Sec.\ \ref{sec:conf}.

In this way, the function $T'$ is determined at $\mathcal C$. To proceed with
the calculation in the $T$-way, we have to extend the function to the outside
of $\mathcal C$. As it was explained above, the particular value of the
extension does not have any physical meaning and can be chosen just by
convenience. A suitable choice is (\ref{24;1}) everywhere (ie., $T'$
independent of $P$). Then, the transversal surface $\mathcal T$ of Sec.\
\ref{sec:conf} is preserved.

The next step of the $T$-way is the propagation of the functions $P'$, $q'$
and $p'$ by the differential equations
\begin{equation}
  \xi_{T'}P' = -1,\quad \xi_{T'}q' = 0,\quad \xi_{T'}p' = 0
\label{25;1}
\end{equation}
out of $\mathcal C$, where we have the initial conditions
\begin{equation}
 P'\vert_{\mathcal C} = 0,\quad q'\vert_{\mathcal C} = q,
  \quad p'\vert_{\mathcal C} = p.
\label{15;2}
\end{equation}
Then, the required values of the Poisson brackets
\[
  \{T',P'\} = 1,\quad \{q',T'\} = 0,\quad \{p',T'\} = 0
\]
are granted. And, for a similar reason as in Sec.\ \ref{sec:transvers}, all
other Poisson brackets will also have the desired values.

From Eq.\ (\ref{24;1}), we easily obtain
\[
  \xi_{T'} = -\sqrt{q^2+p^2}\cos T\frac{\partial}{\partial P} + \frac{p\sin
    T}{\sqrt{q^2+p^2}}\frac{\partial}{\partial q}  
- \frac{q\sin T}{\sqrt{q^2+p^2}}\frac{\partial}{\partial p}.
\]
Eqs.\ (\ref{25;1}) can be solved by the method of characteristics. The
characteristic equations read
\begin{alignat}{2}
  \frac{\partial T}{\partial P'} & = 0, & \qquad \frac{\partial P}{\partial P'}
  & = \sqrt{q^2+p^2}\cos T ,
\label{26;1} \\
  \frac{\partial q}{\partial P'} & = -\frac{p\sin
    T}{\sqrt{q^2+p^2}}, & \qquad \frac{\partial
    p}{\partial P'} & = \frac{q\sin T}{\sqrt{q^2+p^2}}.
\label{26;2}
\end{alignat}
We have already used the fact that the parameter of the characteristic curves
can be chosen to be $-P'$. This follows from the first equation of
(\ref{25;1}). 

We can see immediately that $T$ is an integral of the system and one verifies
easily that $\sqrt{q^2+p^2}$ is another one. Then, the integration of the
system is straightforward and we obtain
\begin{alignat}{2}
  T & = T_0, & \qquad P & = P'\sqrt{q_0^2+p_0^2}\cos T_0 ,
\label{a}   \\
  q & = q_0 \cos (\nu_0 P') - p_0 \sin (\nu_0 P'), & \qquad
  p & = q_0 \sin (\nu_0 P') + p_0 \cos (\nu_0 P'), 
\label{b}
\end{alignat}
where
\[
  \nu_0 = \frac{\sin T_0}{\sqrt{q_0^2+p_0^2}}
\]
and we have used the fact that $P = P' =0$ at $\mathcal C$. The integration
constants $T_0$, $q_0$ and $p_0$ are the values of the coordinates $T$, $q$,
and $p$ at the point where the characteristic intersects the constraint
surface $\mathcal C$.

Everywhere along the characteristic passing through the point $(T_0,q_0,p_0)$,
the functions $q'$ and $p'$ must have the values
\begin{equation}
  q' =  q_0,\quad p' = p_0;
\label{27;1}
\end{equation}
This is a consequence of Eq.\ (\ref{25;1}). The function $T'$ is constant
along each characteristic because of the trivial equation $\xi_{T'}T' = 0$.
Hence, the value of $T'$ along the characteristic (\ref{a}) and (\ref{b}) is
\begin{equation}
  T' = \sqrt{q^{\prime 2}+p^{\prime 2}}\sin T_0.
\label{28;1}
\end{equation}
If we substitute Eqs.\ (\ref{27;1}) and (\ref{28;1}) into Eqs.\ (\ref{a}) and
(\ref{b}), we obtain
\begin{eqnarray}
 P & = & P'\sqrt{q^{\prime 2}+p^{\prime 2} - T^{\prime 2}},
\label{28;2}  \\
 q & = & q'\cos(\nu'P') - p'\sin(\nu'P'),
\label{28;3} \\
 p & = & q'\sin(\nu'P') + p'\cos(\nu'P'),
\label{28;4}
\end{eqnarray}
where
\[
  \nu' := \frac{T'}{q^{\prime 2}+p^{\prime 2}}.
\]
Eqs.\ (\ref{28;3}), (\ref{28;4}) and (\ref{24;1}) yield
\begin{equation}
  T = \arcsin\frac{T'}{\sqrt{q^{\prime 2}+p^{\prime 2}}}.
\label{28;5}
\end{equation}
Eqs.\ (\ref{28;2})--(\ref{28;5}) are the desired transformation formulas
between the two coordinate systems. From the construction, it follows that the
transformation is canonical; this can be verified by direct calculation.

The inverse transformation is given by Eq.\ (\ref{24;1}) together with
\begin{eqnarray}
 P' & = & \frac{P}{\sqrt{q^2+p^2}\cos T},
\label{29;1}  \\
 q' & = & q\cos(\nu P) + p\sin(\nu P),
\label{29;2} \\
 p' & = & -q\sin(\nu P) + p\cos(\nu P),
\label{29;3}
\end{eqnarray}
where
\[
  \nu := \frac{\tan T}{q^2+p^2}.
\]
The transformation functions (\ref{24;1}) and (\ref{29;1})--(\ref{29;3}) are
differentiable everywhere inside $\mathcal P$, that is for the values of the
coordinates $T$, $P$, $q$ and $p$ within the bounds (\ref{7;3}). The
corresponding ranges of the coordinates $T'$, $P'$, $q'$ and $p'$ are
\begin{align*}
  (q',p') & \in {\mathbf R}\setminus \{0\}, \\ 
  T' & \in \left(-\sqrt{q^{\prime 2} + p^{\prime 2}}, 
    \sqrt{q^{\prime 2} + p^{\prime 2}}\right),\\ 
  P' & \in
  \left(-\infty,\frac{1}{2}\sqrt{\frac{q^{\prime 2} + p^{\prime 2}}{q^{\prime
  2} + p^{\prime 2} - T^{\prime 2}}}\right).
\end{align*}
Finally, we observe that the transformation is {\em not} of the form
(\ref{20;1}) and (\ref{20;2}).

\subsubsection{CMC time}
\label{sec:CMC}
The external mean curvature $L$ of the surfaces $t =$ const has the
following value for the metric (\ref{1;1}):
\[
  L = -\frac{1}{3a}\frac{da}{N dt}.
\]
For the conformal time $T$, $N = a$, and we obtain from equation
(\ref{7;1}) 
\[
  L = \frac{1}{3\sqrt{q^2+p^2-2P}}\ \frac{\sin T}{\cos^2T}.
\]

In this section, we shall choose the time function $T''$ to be equal to $L$ at
the constraint surface $P = 0$. We call this coordinate {\em CMC time}
(constant mean curvature). Again, we shall extend this function to the whole
of $\mathcal P$ so that it is independent of $P$:
\[
  T'' = \frac{1}{3\sqrt{q^2+p^2}}\ \frac{\sin T}{\cos^2T}.
\]
The same method as in Sec.\ \ref{sec:proper} leads to the transformation
formulae
\begin{eqnarray*}
 P'' & = & 3\sqrt{q^2+p^2}\frac{\cos^3T}{1+\sin^2T}\ P, \\
 q'' & = & q\cos(\tilde{\nu} P) + p\sin(\tilde{\nu} P), \\
 p'' & = & -q\sin(\tilde{\nu} P) + p\cos(\tilde{\nu} P),
\end{eqnarray*}
where
\[
  \tilde{\nu} := \frac{1}{q^2+p^2}\ \frac{\sin T\cos T}{1 + \sin^2T}.
\]

The transformation is again differentiable everywhere in $\mathcal P$. The
range of the coordinate $T''$ in $\mathcal P$ is the whole real axis, those of
$q''$ and $p''$ remain the same as those of $q$ and $p$, and the range of
$P''$ can be described by its boundary, defined parametrically as follows:
\begin{align*}
  P''_{\text{boundary}} & =
  \frac{3}{2}[(q'')^2+(p'')^2]^{3/2}\frac{\cos^3T}{1+\sin^2T}, \\
  T''_{\text{boundary}} & =  \frac{1}{3\sqrt{(q'')^2+(p'')^2}}\ \frac{\sin
  T}{\cos^2T},
\end{align*}
where $T \in (-\pi/2,+\pi/2)$.

\section{Torus sector of 2+1 gravity}
\label{sec:torus}
Our second model is the partially reduced torus sector of the 2+1 gravity
without sources and with zero cosmological constant. We shall use the form of
the metric
\begin{equation}
  ds^2 = -N^2dt^2 + e^{q^3-q^1}(du^1)^2 + e^{q^3+q^1}\left(du^2 +
  q^2du^1\right)^2 
\label{1,1}
\end{equation}
and the action
\begin{equation}
  S = \int dt(p_i\dot{q}^i - N{\mathcal H}),
\label{1,2}
\end{equation}
where
\begin{equation}
  {\mathcal H} = \frac{1}{2}\,e^{-q^3}\left(p_3^2 - p_1^2 -
  e^{-2q^1}p_2^2\right),
\label{1,3}
\end{equation}
as written down by Moncrief \cite{M1}. Here, $t$, $u^1$ and $u^2$ are
coordinates on the three-dimensional spacetime of topology ${\mathbf R}\times
S^1\times S^1$ chosen such that $t=$ const are the CMC surfaces and $u^A \in
(0,2\pi)$ for $A = 1,2$ are coordinates on the torus such that $x^A =$ const
are closed geodesics of the space metric. Such coordinates can always be
chosen \cite{M1}; using this ``spatially homogeneous gauge'', Moncrief reduced
the field model to a mechanical model with a finite number of degrees of
freedom. These are represented by the Teichm\"{u}ller parameters $q^1$ and
$q^2$. The coordinate $q^3$ is related to the surface area $\mathcal F$ of the
$T =$ const surface by the formula
\[
  {\mathcal F} = 4\pi^2e^{q^3}.
\]

This model is very interesting because it admits solutions with higher
symmetry. All solution spacetimes are spatially homogeneous, invariant with
respect to the abelian group $(u^1,u^2) \mapsto (u^1 + \Delta u^1,u^2 + \Delta
u^2)$; the time evolution of the tori leads to expansion or
contraction. However, if $p_1 = p_2 = 0$, then also $p_3 = 0$ and we obtain
static tori. This is an additional symmetry. Observe that the constraint
surface defined by ${\mathcal H} = 0$ has a conical singularity at these
points. 

Our aim is to find out if the Kucha\v{r} description can incorporate the
points of higher symmetry. The strategy will be to transform the model to the
Kucha\v{r} variables everywhere except at the points of higher symmetry.
There, the transformation becomes singular. We shall try to extend the
resulting Kucha\v{r} description. If we manage, then the extended Kucha\v{r}
description will not be equivalent to the original Moncrief one because the
transformation between them is singular at the points with symmetry.

\subsection{The constraint surface}
\label{sec:constr}
The phase space $\mathcal P$ of the model with the action (\ref{1,2}) is
${\mathbf R}^6$ and the canonical chart $(q^1,q^2,q^3,p_1,p_2,p_3)$ covers the
whole manifold. The symplectic form is $\Omega = d\Theta$, where the Liouville
form $\Theta$ reads
\[
  \Theta = p_idq^i,\quad i = 1,2,3.
\]

The constraint surface $\mathcal C$ is defined by the constraint function
$\mathcal H$ of Eq.\ (\ref{1,3}). Its manifold structure is ${\mathbf R}^3
\times C$, where ${\mathbf R}^3$ is covered by the coordinates $q^1$, $q^2$
and $q^3$, and $C$ is a 2-cone. The tip $\mathcal S$ of the cone $p_1 = p_2 =
p_3 = 0$ is a three-dimensional surface. At the points of $\mathcal S$,
$\mathcal H$ and the gradient of $\mathcal H$ both vanish. The canonical
transformation generated by the function $\mathcal H$ is trivial at $\mathcal
S$. This corresponds to the triviality of evolution of initial data in static
toroidal spacetimes.  Thus, each point of $\mathcal S$ is a whole trajectory
of the Hamiltonian action. At all other points of $\mathcal C$, grad$\mathcal
H$ is non-vanishing and so the trajectories are one-dimensional.

The constraint manifold is an embedded hyper-surface locally, at each point of
${\mathcal C} \setminus {\mathcal S}$. There, we have a differential structure
and the pull-back $\Omega_{\mathcal C}$ of $\Omega$ to $\mathcal C$.
$\Omega_{\mathcal C}$ is only a pre-symplectic form because it is degenerate.
At the points of $\mathcal S$, no such structure is well-defined. 

The sub-manifold ${\mathcal C} \setminus {\mathcal S}$ is the constraint
surface of the truncated model; it consists of two components. As extensions
of these two parts of the truncated model, we introduce two subsets,
${\mathcal C}^+$ and ${\mathcal C}^-$, of $\mathcal C$; all points of
${\mathcal C}^+$ (${\mathcal C}^-$) satisfy the inequality $p_3 \geq 0$ ($p_3
\leq 0$). ${\mathcal C}^+$ and ${\mathcal C}^-$ are topological ($C^0$)
surfaces. The maps $\varphi_\pm : {\mathbf R}^5 \mapsto {\mathcal C}^\pm$
defined by
\[
  \varphi_\pm(x^1,x^2,x^3,y_1,y_2) = (q^1,q^2,q^3,p_1,p_2)
\]
such that $q^i = x^i$ for all $i = 1,2,3$, $p_1 = y_1$, $p_2 = y_2$ and $p_3 =
\pm\sqrt{y_1^2 + e^{-2x^1}y_2^2}$ are both homeomorphisms. They are not
differentiable at $y_1 = y_2 = 0$. Hence, ${\mathcal C}^\pm$ are topological
surfaces with conical singularities at $y_1 = y_2 = 0$. The constraint set
$\mathcal C$ is, however, more singular than that. It has a bifurcation at
$\mathcal S$, where both ${\mathcal C}^+$ and ${\mathcal C}^-$ coincide. The
pre-symplectic form $\Omega_{\mathcal C}$ on ${\mathcal C}^\pm \setminus
{\mathcal S}$ is simply $dy_1\wedge dx^1 + dy_2\wedge dx^2$.

The bifurcation is connected with the way in which the time reversal acts on
the ADM variables. Let us first do a few general remarks concerning the time
reversal. The trajectories at the constraint surface can be considered as
classes of an equivalence relation \cite{FMa}; two initial data are equivalent
if they evolve into maximal solutions that are isometric to each other. We
have, however, to restrict this isometry to the component of unity of the
diffeomorphism group. In particular, it has to preserve all orientations.
Thus, two data from different trajectories can still evolve to isometric
spacetimes, but they must then have different orientations.

For our models, only time orientation exists.  We observe that the map
${\mathbf T}(q^i,p_i) = (q^i,-p_i)$ is anti-symplectic, takes ${\mathcal C}^+$
into ${\mathcal C}^-$ and vice versa. ${\mathbf T}$ coincides with the change
of initial data that is brought about by the time reversal in the solution
spacetimes. The time reversal maps eg.\ an expanding spacetime onto a
contracting one. ${\mathbf T}$ has a well-defined projection ${\mathbf T}_p$
to the physical phase space. ${\mathbf T}_p$ is trivial at ${\mathcal S}$. The
reason is that the two possible time orientations of a static spacetime cannot
be distinguished by their ADM data. We have, therefore, some motivation to
consider all points of ${\mathcal C}^+$ (non-contracting spacetimes) as
physically different from all points of ${\mathcal C}^-$ (non-expanding
spacetimes). This will remove the bifurcation an will leave us just with the
conical singularities.

\subsection{The physical phase space}
\label{sec:physph}
We first construct the two components $\Gamma_\pm$ of the truncated physical
phase space corresponding to ${\mathcal C}^\pm \setminus {\mathcal S}$. The
truncated space is defined as the quotient manifold $\Gamma_\pm :=
({\mathcal C}^\pm \setminus {\mathcal S})/$trajectories. Let us calculate the
trajectories.

To integrate the canonical equations that are implied by the action
(\ref{1,2}), we choose a particular gauge. This gauge will be useful for 
other aims, too. The value of the associated lapse function is:
\begin{equation}
  N = e^{q^3}.
\label{11,1}
\end{equation}
The corresponding time coordinate $T$ has the following relation to the
proper time $\tau$ along the solution spacetimes
\begin{equation}
  e^{q^3}dT = d\tau.
\label{dtau}
\end{equation}
The canonical equations for the Hamiltonian 
\begin{equation}
  P = N{\mathcal H} = \frac{1}{2}(p_3^2 - p_1^2 - e^{-2q^1}p_2^2)
\label{12,7}
\end{equation}
can be written in the following form
\begin{alignat}{2}
  \dot{q}^1 & = -p_1, & \qquad \dot{p}_1 & = -e^{-2q^1}p_2^2, 
\label{12,1} \\
  \dot{q}^2 & = -e^{-2q^1}p_2, & \qquad \dot{p}_2 & = 0,
\label{12,2} \\
  \dot{q}^3 & = p_3, & \qquad \dot{p}_3 & = 0.
\label{12,3} 
\end{alignat}
At the constraint surface, $P = 0$, but $P$ is an integral of these equations
everywhere in $\mathcal P$. Second Eq.\ (\ref{12,3}) implies then that
\begin{equation}
  K := \sqrt{p_1^2 + e^{-2q^1}p_2^2}
\label{12,4}
\end{equation}
is also an integral. A straightforward but lengthy calculation gives a general
solution to Eqs.\ (\ref{12,1})--({\ref{12,3}) everywhere in $\mathcal P$:
\begin{align}
  q^1 & = q^1_0 + \ln\left(\frac{K-p_1^0}{2K}e^{KT} +
  \frac{K+p_1^0}{2K}e^{-KT}\right),
\label{12,5} \\
  p_1 & = -K\frac{(K-p_1^0)e^{KT} - (K+p_1^0)e^{-KT}}{(K-p_1^0)e^{KT} +
  (K+p_1^0)e^{-KT}}, 
\label{12,6} \\
  q^2 & = q^2_0 - e^{-2q^1}p_2^0\frac{e^{KT} - e^{-KT}}{(K-p_1^0)e^{KT} +
  (K+p_1^0)e^{-KT}}, 
\label{13,1} \\
  p_2 & = p_2^0,
\label{13,2} \\
  q^3 & = p_3^0T + q^3_0,
\label{13,3a} \\
  p_3 & = p_3^0;
\label{13,3b}
\end{align}
this solution runs through the point $(q_0^i,p_i^0)$ for $T = 0$. At
${\mathcal C}^\pm$, we have
\[
  p_3^0 = \pm K_0,
\]
where
\[
  K_0 := \sqrt{(p_1^0)^2 + e^{-2q_0^1}(p_2^0)^2}.
\]
The subset ${\mathcal S}^\pm$ of ${\mathcal C}^\pm$ is defined by $K = 0$.
Then Eqs.\ (\ref{12,5})--(\ref{13,3b}) become
\[
  q^i = q_0^i, \quad p_i = p_i^0.
\]

The range of the time coordinate $T$ is $(-\infty,\infty)$. Eqs.\ (\ref{dtau})
and (\ref{13,3a}) show, however, that $q^3 \rightarrow -\infty$ is a
singularity: it is reached in a finite proper time. We obtain easily from Eq.\
(\ref{dtau})
\[
  \tau_{\text{sing}} - \tau_0 = -\frac{e^{q^3}}{p_3},
\]
where $\tau_0$ is the value of proper time at the point $T = 0$.

An important property of Eqs.\ (\ref{12,7})--(\ref{12,3}) is the so-called
{\em linearization instability} \cite{FMM}, \cite{M3} at the points of
$\mathcal S$, where $p_1 = p_2 = p_3 = 0$. If we expand these equations around
the static solutions, the constraint
\[
  p_3^2 - p_1^2 - e^{-2q^1}p_2^2 = 0
\]
becomes trivial, $0 = 0$, in the first order. The first non-trivial
contribution to it is the second-order one,
\[
  (\delta_1p_3)^2 - (\delta_1p_1)^2 - e^{-2q_0^1}(\delta_1p_2)^2 = 0.
\]
This equation does not, however, contain any second order correction
$\delta_2q^i$ and $\delta_2p_i$. It is a second order condition for the first
order corrections $\delta_1q^i$ and $\delta_1p_i$. Thus, some solutions of the
first order equations (``linearized equations'') are spurious.

In the set ${\mathcal C}^\pm \setminus {\mathcal S}$, the integral $K$ is
positive. Eq.\ (\ref{13,3a}) then implies that $q^3$ is a strictly increasing
function of $T$ on ${\mathcal C}^+ \setminus {\mathcal S}$ and well-defined
for $T \in (-\infty,\infty)$. The range of the function is again
$(-\infty,\infty)$. Similarly, $q^3$ is strictly decreasing on ${\mathcal C}^-
\setminus {\mathcal S}$. Hence, for both cases, the surface ${\mathcal T}^\pm$
defined by $q^3 = q^3_0$ intersect {\em each} trajectory exactly {\em once},
in a {\em transversal} direction. It is, therefore, a transversal surface for
any value of $q^3_0 \in (-\infty,\infty)$.  The transversal surface can be
described as the following embedding of the manifold ${\mathbf R}^2 \times
({\mathbf R}^2 \setminus \{0\})$ with coordinates $x^1$, $x^2$, $y_1$ and
$y_2$, where $(x^1,x^2) \in {\mathbf R}^2$ and $(y^1,y^2) \in {\mathbf R}^2
\setminus \{0\}$, into ${\mathcal C}_\pm$:
\begin{equation}
  q^1 = x^1,\quad q^2 = x^2,\quad p_1 = y_1,\quad p_2 = y_2,
\label{14,1}
\end{equation} 
\begin{equation}
  q^3 = q^3_0,\quad p_3 = \pm\sqrt{y_1^2 + e^{-2x^1}y_2^2}.
\label{14,2}
\end{equation} 
The pull-back $\omega_\pm$ of the pre-symplectic form $\Omega_{\mathcal C}$ to
$\Gamma_\pm$,
\begin{equation}
  \omega_\pm = dy_1\wedge dx^1 + dy_2\wedge dx^2,
\label{14,3}
\end{equation}
is non-degenerate.

Let us consider two such sections, ${\mathcal T}^\pm_1$ and ${\mathcal
  T}^\pm_2$, defined by $q^3 = q^3_1$ and $q^3 = q^3_2$, respectively. Let the
coordinates on these sections analogous to those defined by Eqs. (\ref{14,1})
and (\ref{14,2}) be $x^1_1$, $x^2_1$, $y_1^1$, $y_2^1$, and $x^1_2$, $x^2_2$,
$y_1^2$, $y_2^2$, respectively. Then at each point of ${\mathcal T}^\pm_1$ a
unique trajectory starts and it intersects ${\mathcal T}^\pm_2$. This defines
a point of ${\mathcal T}^\pm_2$ for each point of ${\mathcal T}^\pm_1$, and we
obtain a map $\phi^\pm_{q^3_1q^3_2}$ between ${\mathcal T}^\pm_1$ and
${\mathcal T}^\pm_2$. We easily find the map from the solution
(\ref{12,5})--(\ref{13,3b}) in terms of coordinates:
\begin{align}
  x^1_2 & = x^1_1 + \ln\left(\frac{K_1-y_1^1}{2K_1}e^{\pm\Delta q^3} +
  \frac{K_1+y_1^1}{2K_1}e^{\mp\Delta q^3}\right), 
\label{15,1} \\
  y_1^2 & = -K_1\frac{(K_1-y_1^1)e^{\pm\Delta q^3} - (K_1+y_1^1)e^{\mp\Delta
  q^3}} {(K_1-y_1^1)e^{\pm\Delta q^3} + (K_1+y_1^1)e^{\mp\Delta q^3}},
\label{15,2} \\
  x_2^2 & = x_1^2 - e^{-2q^1_1}y_2^1 \frac{e^{\pm\Delta q^3} - e^{\mp\Delta
  q^3}}{(K_1-y_1^1)e^{\pm\Delta q^3} + (K_1+y_1^1)e^{\mp\Delta q^3}},
\label{16,1} \\
  y_2^2 & = y_2^1,
\label{16,2}
\end{align}
where $\Delta q^3 := q^3_2 - q^3_1$ and $K_1 := \sqrt{(y_1^1)^2 +
  e^{-2x^1_1}(y_2^1)^2}$. The map $\phi^\pm_{q^3_1q^3_2}$ is, of course, a
symplectic diffeomorphism (as one can also verify by a direct calculation).
The truncated physical phase space $\Gamma_+ \cup \Gamma_-$ can be considered
as the set of all transversal surfaces, all points of which are identified by
the maps analogous to $\phi^\pm_{q^3_1q^3_2}$.

Our next aim is to extend Kucha\v{r} description to the solution spacetimes
with additional symmetry. The corresponding separation between the physical
and the gauge degrees of freedom requires a well-defined physical phase space
as a necessary ingredient. Our first step must, therefore, be an extension of
the truncated spaces $\Gamma_+$ and $\Gamma_-$ to the points of ${\mathcal
  S}^+$ and ${\mathcal S}^-$.

Let us consider the quotient {\em sets} ${\mathcal C}^\pm/$trajectories and
denote the corresponding projection maps by $\pi_\pm$. The quotient topology
is defined as the finest one on ${\mathcal C}^\pm/$trajectories that makes
$\pi_\pm$ continuous. Hence, $\pi_\pm$ is open. Let us denote the resulting
topological spaces by $\bar{\Gamma}_\pm$. They are paracompact, locally
compact, but not Hausdorff. The real problem is, however, that the topological
spaces $\bar{\Gamma}_\pm$ do not carry any natural differential structure.

It is, however, possible to introduce a differentiable structure on
$\bar{\Gamma}_\pm$ that is inherited directly from $\mathcal P$. An atlas for
$\bar{\Gamma}_\pm$ can be defined by transversal surfaces as follows.  Let us
extend each transversal surface ${\mathcal T}^\pm_\kappa$, defined by $q^3 =
\kappa$, to $\bar{\mathcal T}^\pm_\kappa$ by adding all points of ${\mathcal
  S}^\pm$ that satisfy the same equation. This is the two-dimensional subset
$(q^1,q^2) \in {\mathbf R}^2$, $q^3 = \kappa$. The coordinates on
$\bar{\mathcal T}^\pm_\kappa$ can be chosen as $(x^1,x^2,y_1,y_2) \in {\mathbf
  R}^4$ and the embedding formulae coincide with Eqs.\ (\ref{14,1}) and
(\ref{14,2}). The sets $\bar{\mathcal T}^\pm_\kappa$ are topological
sub-manifolds of ${\mathcal C}_\pm$ and {\em differentiable sub-manifolds} of
$\mathcal P$. The symplectic form $\omega^\pm_\kappa$ given by Eq.\ 
(\ref{14,3}) is uniquely extensible to $\bar{\mathcal T}^\pm_\kappa$ by
continuity. 

For a given $\kappa$, each point of $\bar{\mathcal T}^\pm_\kappa$ represents a
unique trajectory, but all points of $\bar{\mathcal T}^\pm_\kappa$ do not
represent all trajectories. Those points of ${\mathcal S}^\pm$ that do not
satisfy the equation $q^3 = \kappa$ are trajectories that do not intersect
$\bar{\mathcal T}^\pm_\kappa$. Hence, to represent all trajectories, we need
{\em all} transversal surfaces, $\kappa \in (-\infty,\infty)$. The points of
$\bar{\mathcal T}^\pm_\kappa$ that do not lie in ${\mathcal S}^\pm$ represent
trajectories that intersect all other transversal surfaces. Hence, to
represent each trajectory by just one point, we have to identify the
transversal surfaces by the maps $\phi^\pm_{\kappa_1\kappa_2}$. The surfaces
$\bar{\mathcal T}^\pm_\kappa$ for all $\kappa \in (-\infty,\infty)$, together
with the maps $\phi^\pm_{\kappa_1\kappa_2}$ form the desired atlas, which we
denote by $\mathcal A$.

As it is usual for manifolds, its topology can be defined by a basis that is a
union of the bases for each chart. Thus, the atlas $\mathcal A$ also defines a
topology on $\bar{\Gamma}_\pm$---let us call it $\mathcal A$-topology. However
the $\mathcal A$-topology and the quotient one do not coincide. For example,
the $\mathcal A$-topology is not {\em paracompact}. To cover
$\bar{\Gamma}_\pm$, we need an uncountable set of charts.  Then the basis of
the topology of $\bar{\Gamma}_\pm$ is not countable and $\bar{\Gamma}_\pm$ is
not paracompact.  The $\mathcal A$-topology also fails to describe the
``nearness'' between different $\kappa$-levels of ${\mathcal S}^\pm$ properly.
Indeed, each $\kappa$-level of ${\mathcal S}^\pm$ is contained in a different
chart, $\bar{\mathcal T}^\pm_\kappa$. ${\mathcal S}^\pm \cap \bar{\mathcal
  T}^\pm_\kappa$ is, therefore, contained in an open set that does not
intersect any other $\kappa$-level. Then, a sequence of points of ${\mathcal
  S}^\pm$ that do not lie in ${\mathcal S}^\pm \cap \bar{\mathcal
  T}^\pm_\kappa$ can never converge to any point of ${\mathcal S}^\pm \cap
\bar{\mathcal T}^\pm_\kappa$ in the $\mathcal A$-topology.

To see that $\bar{\Gamma}_\pm$ is not Hausdorff (in the $\mathcal A$-topology
as well as in the quotient one), let us consider two transversal surfaces
$\bar{\mathcal T}^\pm_1$ and $\bar{\mathcal T}^\pm_2$ defined by $q^3 = q^3_1$
and $q^3 = q^3_2$, respectively. Let $\{Q^\pm_{1n}\} \subset \bar{\mathcal
  T}^\pm_1$ be a point sequence in $\bar{\mathcal T}^\pm_1$ with coordinates
\[
  Q^\pm_{1n} := (x^1_1,x^2_1,y_{1n}^1,y_{2n}^1).
\]
Let $y_{1n}^1 \neq 0$ and $y_{2n}^1 \neq 0$ for all positive integers $n$ and 
\[
  \lim_{n\rightarrow\infty}y_{1n}^1 = 0,\quad
  \lim_{n\rightarrow\infty}y_{2n}^1 = 0. 
\]
This sequence converges in $\bar{\mathcal T}^\pm_1$ to the point 
\[
  Q_1^\pm = (x^1_1,x^2_1,0,0) \in {\mathcal S}_\pm.
\]
All point of the sequence lie outside ${\mathcal S}^\pm$ and so can be
identified with points $Q^\pm_{2n}$ of $\bar{\mathcal T}^\pm_2$ that are
determined by Eqs.\ (\ref{15,1})--(\ref{16,2}). Their coordinates are
\begin{align*}
  x^1_{2n} & = x^1_1 + \ln(\cosh\Delta q^3 - \sin\alpha_n\sinh\Delta q^3), \\
  x^2_{2n} & = x^2_1 - \frac{\cos\alpha_n\sinh\Delta q^3}{\cosh\Delta q^3 -
  \sin\alpha_n\sinh\Delta q^3}, \\
  y^2_{1n} & = \pm K^1_n \frac{-\sinh\Delta q^3 + \sin\alpha_n\cosh\Delta
  q^3}{\cosh\Delta q^3 - \sin\alpha_n\sinh\Delta q^3}, \\
  y^2_{2n} & = y^1_{2n},
\end{align*}
where $\alpha_n$ is defined by
\[
  \sin\alpha_n = \frac{y^1_{1n}}{K^1_n},\quad \cos\alpha_n =
  \frac{y^1_{2n}e^{-x^1_1}}{K^1_n}. 
\]
and
\[
  K^1_n := \sqrt{(y^1_{1n})^2 + e^{-2x_{1n}^1}(y^1_{2n})^2}.
\]
The sequence $\{Q^\pm_{2n}\}$ converges to the point $Q^\pm_{2\alpha} \in
{\mathcal S}^\pm$ that is given by $(x^1_2,x^2_2,0,0)$ if and only if
$\lim_{n\rightarrow\infty}\alpha_n = \alpha$ exists. Then
\begin{align}
  x^1_2 =  & = x^1_1 + \ln(\cosh\Delta q^3 - \sin\alpha\sinh\Delta q^3),
\label{x1} \\
  x^2_{2n} & = x^2_1 - \frac{\cos\alpha\sinh\Delta q^3}{\cosh\Delta q^3 -
  \sin\alpha\sinh\Delta q^3}.
\label{x2}
\end{align}
Each sequence $\{Q^\pm_{2n}\}$ with a converging $\alpha_n$ has then a unique
limit in $\bar{\mathcal T}^\pm_2$, the possible limit points of all converging
sequences fill out a closed curve in ${\mathcal S}^\pm \cap {\mathcal
  T}^\pm_2$ defined by Eqs.\ (\ref{x1}) and (\ref{x2}) for $\alpha \in S^1$.
Each point $Q^\pm_{2\alpha}$ is different from $Q^\pm_1$ because its
coordinates $q^3$ in ${\mathcal C}^\pm$ differ. Thus, one sequence can have
two different limits and the space cannot be Hausdorff.

An important property of the atlas can easily be shown: it is not unique.
Indeed, we could slightly deform the transversal surfaces $\bar{T}^\pm_\kappa$
so that they remain transversal in ${\mathcal C}^\pm \setminus {\mathcal
  S}^\pm$ and so that their intersections with ${\mathcal S}^\pm$ remain
two-dimensional. Let the new transversal surfaces be defined by some equation
of the form $f(q^1,q^2,q^3,p_1,p_2) = \kappa$, $\kappa \in (-\infty,\infty)$.
The intersection with ${\mathcal S}^\pm$ is given by
\[
  f(q^1,q^2,q^3,0,0) = \kappa
\]
and it will generically intersect the surface $q^3 = \kappa$ in a
curve
\[
  f(q^1,q^2,\kappa,0,0) = \kappa.
\]
Thus, the intersection between some transversal surface of the first family
with some transversal surface of the second family will {\em not} be open. It
follows that we cannot simply add the new family of transversal surfaces to
the old atlas, and so the manifold defined by each of the two atlases will be
different.

Any of these atlases can serve as a basis for a construction of Kucha\v{r}
decomposition. This will be shown in the next section.

\subsection{Transformation to embedding variables}
\label{sec:ev}
To construct the transformation we shall use the $P$-way as described in
Sec.\ \ref{sec:fried}. The Hamiltonian that we choose corresponds to the value
of the lapse function (\ref{11,1}). The Hamiltonian itself has the form
(\ref{12,7}), the canonical equations are (\ref{12,1})--(\ref{12,3}) and the
general solution to these equations is given by Eqs.\
(\ref{12,5})--(\ref{13,3b}). 

The next step is a choice of transversal surface in $\mathcal P$. Our choice
is a straightforward extension of the transversal surfaces $\bar{\mathcal
  T}^\pm_\kappa$ of Sec.\ \ref{sec:physph} to $\mathcal P$ by the equation
$q^3 = \kappa$. Let us denote the result by $\tilde{\mathcal T}^\pm_\kappa$.
It is transversal everywhere in ${\mathcal P} \setminus {\mathcal C}$ as long
as $p_3 \neq 0$ because of Eq.\ (\ref{13,3a}).

Eqs.\ (\ref{12,7}) and (\ref{12,4}) imply that
\[
  p^0_3 =\pm\sqrt{2P + K^2}.
\]
The function $p_3^0$ remains non zero in the part of ${\mathcal P}$ that is
determined by the following inequality
\begin{equation}
  P > -\frac{1}{2}K^2.
\label{ineq}
\end{equation}
The trajectories lying at its boundary have $T$-independent surface area,
$e^{q^3}$, but they are not static if $P \neq 0$: $q^1$ and $q^2$ are evolving
in a non trivial way. At ${\mathcal C}^\pm$, where $P = 0$ (and so $K = 0$
at the boundary), there are no problems because the corresponding trajectories
are just points. However, the trajectories at $P < 0$, $K = \sqrt{-2P}$ are
not points and they are lying in the surfaces $q^3 =$ const. Thus,
$\tilde{\mathcal T}^\pm_\kappa$ ceases to be transversal at this boundary.

It is helpful to realize that $\mathcal C$ divides $\mathcal P$ into three
disjoint parts similarly as the light cone divides Minkowski spacetime into
the future interior, the past interior and the exterior of the light
cone. Thus, we have ${\mathcal P}^+$ defined by $P > 0$ and $p_3 > 0$,
${\mathcal P}^-$ by $P > 0$ and $p_3 < 0$ and ${\mathcal P}^0$ by $P < 0$. The
surface $p_3 = 0$ separates $\mathcal P$ into two halves, one with $p_3 > 0$
and one with $p_3 < 0$. The transversal surfaces $\tilde{\mathcal T}^+_\kappa$
cover ${\mathcal P}^+$, ${\mathcal C}^+$ and the $p_3 > 0$ part of ${\mathcal
  P}^0$. Let us denote this set by $\tilde{\mathcal P}^+$. Similarly, the
surfaces $\tilde{\mathcal T}^-_\kappa$ cover ${\mathcal P}^-$, ${\mathcal
  C}^-$ and the $p_3 < 0$ part of ${\mathcal P}^0$; this set will be denoted
by $\tilde{\mathcal P}^-$. Observe that $\tilde{\mathcal P}^\pm$ is not an
open subset of $\mathcal P$; it has the boundary ${\mathcal S}^\pm$. At all
points of ${\mathcal C}^\pm \setminus {\mathcal S}^\pm$, the surfaces
$\tilde{\mathcal T}^\pm_\kappa$ are well-defined at both sides of ${\mathcal
  C}^\pm$. At ${\mathcal S}^\pm$, they are defined only in the ${\mathcal
  P}^\pm$ side of the surface ${\mathcal C}^\pm$.

For each $\kappa$, the solution (\ref{12,5})--(\ref{13,3b}) with $q^3_0 =
\kappa$ and $p_3^0 = \pm\sqrt{2P + K^2}$ cover certain part $\tilde{\mathcal
  P}^\pm_\kappa$ of $\tilde{\mathcal P}^\pm$. We are going to use Eqs.\ 
(\ref{12,5})--(\ref{13,3b}) to define maps from $\tilde{\mathcal T}^\pm_\kappa
\times {\mathbf R}$ into $\tilde{\mathcal P}^\pm_\kappa$ that we call
$\psi^\pm_\kappa$. Let the coordinates on $\tilde{\mathcal T}^\pm_\kappa$ be
$x^1$, $x^2$, $y_1$, $y_2$, $P$ and that on $\mathbf R$ be $T$. Eqs.\ 
(\ref{12,5})--(\ref{13,3b}) have to be rewritten, to define $\psi^\pm_\kappa$,
in such a way that $q^3_0= \kappa$, $q^1_0 = x^1$, $q^2_0 = x^2$, $p_1^0 =
y_1$ and $p_2^0 = y_2$:
\begin{align}
  q^1 & = x^1 + \ln\left(\frac{Y-y_1}{2Y}e^{YT} +
  \frac{Y+y_1}{2Y}e^{-YT}\right),
\label{5,1} \\
  p_1 & = -Y\frac{(Y-y_1)e^{YT} - (Y+y_1)e^{-YT}}{(Y-y_1)e^{YT} +
  (Y+y_1)e^{-YT}}, 
\label{6,1} \\
  q^2 & = x^2 - e^{-2x^1}y_2\frac{e^{YT} - e^{-YT}}{(Y-y_1)e^{YT} +
  (Y+y_1)e^{-YT}}, 
\label{6,2} \\
  p_2 & = y_2,
\label{6,3} \\
  q^3 & = \pm T\sqrt{2P + Y^2} + \kappa,
\label{6,4a} \\
  p_3 & = \pm\sqrt{2P + Y^2},
\label{6,4b}
\end{align}
where
\begin{equation}
  Y = \sqrt{y_1^2 + e^{-2x^1}y_2^2}.
\label{6,5}
\end{equation}
The map $\psi^\pm_\kappa$ is invertible on $\tilde{\mathcal P}^\pm_\kappa
\setminus {\mathcal S}^\pm_\kappa$ where $p_3 \neq 0$. The inverse
transformation is described by the following equations:
\begin{align}
  T & = \frac{q^3-\kappa}{p_3},
\label{23,1} \\
  P & = \frac{1}{2}\left(p_3^2 - p_1^2 - e^{-2q^1}p_2^2\right), 
\label{23,2} \\
  x^1 & = q^1 + \ln\frac{(K+p_1)e^{KT} + (K-p_1)e^{-KT}}{2K}, 
\label{23,3} \\
  y_1 & = K\frac{(K+p_1)e^{KT} - (K-p_1)e^{-KT}}{(K+p_1)e^{KT} +
(K-p_1)e^{-KT}}, 
\label{24,1} \\
  x^2 & = q^2 - e^{-2q^1}p_2\frac{e^{KT} - e^{-KT}}{(K+p_1)e^{KT} +
(K-p_1)e^{-KT}}, 
\label{24,2} \\
  y_2 & = p_2,
\label{24,3}
\end{align}
where $K$ is defined by Eq.\ (\ref{12,4}) and the substitution
(\ref{23,1}) is to be made for $T$ in Eqs.\
(\ref{23,3})--(\ref{24,2}).

The functions $T$ and $P$ given by Eqs.\ (\ref{23,1}) and (\ref{23,2})
are singular at ${\mathcal S}^\pm$ where $p_3 = p_1 =p_2 = 0$:
\begin{align*}
  dT & = \frac{1}{p_3}\,dq^3 - \frac{q^3-\kappa}{p_3}\,dp_3, \\
  dP & = e^{-2q^1}p_2^2dq^1 - p_1dp_1 - e^{-2q^1}p_2dp_2 + p_3dp_3.
\end{align*}
$dT$ diverges and $dP$ goes to zero. Still, the pull-back
$\Omega_\kappa$ of the symplectic form $\Omega$ by $\psi^\pm_\kappa$
from ${\mathcal P}^\pm_\kappa \setminus {\mathcal S}^\pm_\kappa$
remains regular at these points. An easy calculation reveals that
\[
  \Omega_\kappa = dP\wedge dT + dy_1\wedge dx^1 + dy_2\wedge dx^2.
\]
Hence, there is a trivial extension of $\Omega_\kappa$ to the whole of
$\tilde{\mathcal T}^\pm_\kappa \times {\mathbf R}$. This seems to be a
general pattern that may hold for all conical singularities.

The range of $\psi^\pm_\kappa$ is not the whole of $\tilde{\mathcal P}^\pm$:
it contains all trajectories of $\tilde{\mathcal P}^\pm \setminus {\mathcal
  S}^\pm$, but it does not contain the point trajectories of ${\mathcal
  S}^\pm$ that do not satisfy the equation $q^3 = \kappa$. To cover the whole
of $\tilde{\mathcal P}^\pm$, we have to use $\psi^\pm_\kappa$ for all $\kappa
\in (-\infty,\infty)$.

Let, for two different $\kappa$'s, $\kappa_1$ and $\kappa_2$, the
corresponding maps be $\psi^\pm_1$ and $\psi^\pm_2$, and let their domains
have coordinates $(x_1^1,x_1^2,y_1^1,y_2^1,P_1,T^1)$ and
$(x_2^1,x_2^2,y_1^2,y_2^2,P_2,T^2)$, respectively. The maps $\psi^\pm_1$ and
$\psi^\pm_2$ are invertible and $C^\infty$ where their ranges overlap, so they
define a map $(\psi^\pm_2)^{-1} \circ \psi^\pm_1$ on
$(\psi^\pm_1)^{-1}({\mathcal P}^\pm_1 \cap {\mathcal P}^\pm_2)$. This map can
be explicitly calculated from Eqs.\ (\ref{5,1})--(\ref{6,4b}) and
(\ref{23,1})--(\ref{24,3}); it turns out to be a $C^\infty$ symplectomorphism.
Hence, the manifold $\tilde{\mathcal P}^\pm_K$ that results by pasting
together all charts ${\mathcal T}^\pm_\kappa \times {\mathbf R}$ by these maps
is a $C^\infty$ symplectic manifold.

$\tilde{\mathcal P}^\pm_K$ is Hausdorff, as any sequence that converges to
some point of $\tilde{\mathcal T}^\pm_1$ in the chart corresponding to
$\kappa_1$ will diverge in the chart corresponding to $\kappa \neq \kappa_1$,
say $\kappa = \kappa_2$. This can be seen from the relation between $T_2$ and
$T_1$ obtained from Eqs.\ (\ref{23,1}) and (\ref{6,4a}):
\[
  T_2 = T_1 \mp \frac{\kappa_2-\kappa_1}{\sqrt{2P_1 + Y_1^2}}.
\]
Observe that $P_1 = P_2$ and so the constraint set ${\mathcal C}^\pm_K$ in
$\tilde{\mathcal P}^\pm_K$ can be described by the single equation
\[
  P = 0.
\]
It is a smooth (Hausdorff) sub-manifold of $\tilde{\mathcal P}^\pm_K$.

${\mathcal C}^\pm_K$ can be considered as a fiber bundle with the basis
$\tilde{\Gamma}^\pm$ and fiber $\mathbf R$. It is defined by the
trivializations $\tilde{\mathcal T}^\pm_\kappa \times {\mathbf R}$ and by
pasting maps of the type $(\psi^\pm_2)^{-1}\circ\psi^\pm_1$ restricted to $P =
0$.

In this way, we have constructed a kind of Kucha\v{r} description for the
torus sector that includes the static tori. The construction is not unique,
and the result is also somewhat strange. The origin of the problem is in the
pathological structure of the physical phase space, which is shared with the
ADM description.

The Kucha\v{r} charts cover only the part $\tilde{\mathcal P}^+ \cup
\tilde{\mathcal P}^-$ of the original phase space $\mathcal P$; the points
of $\mathcal S$ are covered two times. They have to
satisfy the inequalities 
\[
  T \in (-\infty,\infty),\quad (x^1,x^2,y_1,y_2) \in {\mathbf R}^4,
\]
and 
\[
  P \in \left(-\frac{1}{2}(y_1^2 + e^{-2x^1}y_2^2),\infty\right) \cup \{0\}
\]
in each chart.

The new dynamical equations for the variables $T$, $P$, $x^1$, $x^2$, $y_1$ and
$y_2$ in each chart are very simple:
\[
  \dot{x}^1 = \dot{x}^2 = \dot{y}_1 = \dot{y}_1 = \dot{P} = 0,
\]
and 
\[
  \dot{T} = N', \quad P = 0,
\]
where $N'$ is the new Lagrange multiplier that enforces the new constraint.
These equations are manifestly linearization stable (because they are linear).
The deeper reason for the stability is the absence of conical singularities in
the new description.

\section{Concluding remarks}
\label{sec:concl}
We have studied some global properties of the transformation from the ADM to
the Kucha\v{r} description in two minisuperspace models. We have found in
all cases that the Kucha\v{r} description is globally inequivalent to the ADM
description. The solutions to the two corresponding sets of dynamical
equations, however, always completely coincide.

The first interesting feature that we have met are the non-trivial boundaries
for Kucha\v{r} variables. As yet, three different kinds of boundaries have
been detected. First, there are boundaries due to singularities in solutions
of Einstein's equations. It does not seem sensible to propose any general
method of dealing with these singularities in the classical version of the
theory. The hope is that the quantum theory will cure them in some way (for an
example, see \cite{shell}). We just ignore these boundaries.

Second, we have found bounds for the variables conjugate to embeddings, in our
case $P$, in all models. The meaning of the bounds is simply that the function
$P$ does not attain all values on the ADM phase space (Sec.\ \ref{sec:fried}),
or on a suitable part of it (Sec.\ \ref{sec:torus}).  Our standpoit here
simply is that nothing seems to prevent an extension of Kucha\v{r} phase space
to all values of $P$. The dynamics is not changed by this extension. This is
the reason why we did not made any comment when the bounds became too narrow
so that a part of the constraint surface appeared ``bare'' from one side (the
set ${\mathcal S}^\pm$ of the torus sector of 2+1 gravity model). In fact, we
have seen that most claims concerning the structure of the set ${\mathcal P}
\setminus {\mathcal C}$ are either trivial or gauge dependent. It seems,
therefore, that this structure is not relevant to physical properties of the
system (although it can be used for some methodical purposes). This is one
more reason to consider its extensions as harmless.

Finally, there is a boundary for the embedding variables due to non-space-like
character of some embeddings. This type of boundary has not been encountered
here, but it is quite analogous to the $P$-boundary. It seems that an
extension of the Kucha\v{r} description to all values of embeddings may again
be harmless. Let us consider the extension $\bar{\mathcal P}_K$ of the
Kucha\v{r} phase space that contains all embeddings of the space
Emb$(\Sigma,{\mathcal M})$ for each point of the physical phase space $\Gamma$
and all values of the momenta from the space $T^*_X\text{Emb}(\Sigma,{\mathcal
  M})$ for each pair of points of $\Gamma \times \text{Emb}(\Sigma,{\mathcal
  M})$. The inequalities that both the embeddings $X$ and their conjugate
momenta $P$ have to satisfy are then telling us that the ADM phase space
$\mathcal P$ is a proper subset of $\bar{\mathcal P}_K$. Does it mean that the
ADM phase space is too small, or that the Kucha\v{r} phase space is
unnecessarily large? 

There is one argument in favor of the first claim.  Isham and Kucha\v{r}
\cite{IK} have studied the action of the diffeomorphism group in a phase space
of ADM type. They observed that, given any fixed diffeomorphism $\varphi \in
\text{Diff}{\mathcal M}$, there is a Cauchy surface $\Sigma$ in any solution
spacetime such that $\varphi(\Sigma)$ is not space-like. If the ADM variables
associated with the surface $\Sigma$ are $q_{kl}$ and $\pi^{kl}$, then the
representative of $\varphi$ acting on the phase space must map the point
$(q_{kl},\pi^{kl})$ out of the phase space.  They have concluded that only its
Lie algebra but not the group Diff$\mathcal M$ itself has a well-defined
action on the phase space. This can be compared with the situation in the
Yang-Mills field theory, where the full gauge group has a well-defined action
on the phase space. We also easily recognize that Diff$\mathcal M$ acts
without hindrance on $T^*\text{Emb}(\Sigma,{\mathcal M})$ and so it has a
well-defined action on the extended Kucha\v{r} phase space $\bar{\mathcal
  P}_K$.

Other problems that we have studied are connected with the points that
correspond to solutions of higher symmetry. For a model---the torus sector of
2+1 gravity---we have found a description by Kucha\v{r} variables including
such points. The new description is smooth: there is no bifurcation, no
conical singularity and no linearization instability. It is not equivalent
to the ADM description given in \cite{M1}. This point ought to be stressed:
``passing'' to the Kucha\v{r} description is {\em not} just a coordinate
transformation on the phase space. A mere coordinate transformation could not
remove bifurcation, conical singularity or linearization instability.

The essence of the problem with the bifurcation, the conical singularity and
the linearization instability in the case of the ADM description is that the
fields $q_{kl}$ and $\pi^{kl}$ cannot distinguish between two Cauchy surfaces
that are linked by an isometry. Two such Cauchy surfaces then define one and
the same point in the ADM phase space. However, the two Cauchy surfaces can
surely be distinguished from each other by an observer living in the
corresponding solution spacetime. Hence, the ADM description is not true if
there are {\em any} symmetries. This is to be compared with the description by
the embedding variables, which is true, so to speak, by definition. The
conical singularity and the linearization instability are consequences of this
untrue description only if some additional conditions are satisfied. First,
the solutions with the symmetry must form a subset of all solutions that also
include solutions without the symmetry (the static tori are solutions as well
as the expanding and contracting tori are). Second, the symmetry must be
continuous (Killing vectors).

The bifurcation of the ADM constraint surface is caused by an additional
discrete symmetry: the time reversal. The static spacetimes are invariant with
respect to it; the expanding and contracting spacetimes are not. The ADM
description identifies the two possible time orientations of the static
spacetimes, but it always distinguishes the points corresponding to the
contracting from those corresponding to the expanding spacetimes. In the phase
space, the two surfaces, one for the non-contracting and the other for the
non-expanding spacetimes, are then identified at the points corresponding to
the static spacetimes. In this way, the bifurcation of the constraint surface
$\mathcal C$ within the ADM description comes about.

In the case of the torus model, we have also seen that there is no global
gauge at the constraint surface $\mathcal C$. It may be possible to choose one
smooth lapse function $N$ with a domain that includes the whole of $\mathcal
C$, but there is no global {\em transversal surface}. This leads to a
non-trivial fiber-bundle structure for the constraint surface within the
Kucha\v{r} description. Each transversal surface defines a trivialization of
this bundle, but there is no global trivialization. Such a construction has
been considered in \cite{HKi}. In fact, even in the cases that admit a global
gauge, the gauge is not unique. Thus, it is the bundle that represents the
gauge invariant structure of the constraint surface in all cases. Although the
bundle is trivial if a global gauge exists, it does not possess any {\em
  canonical} trivialization. This has been explained in \cite{HKi}.

The present paper focuses on the transformation between the ADM and the
Kucha\v{r} descriptions. This necessarily leads to a comparison of just these
two. One should not, however, forget that there are many other
descriptions. In this respect, it may be interesting to observe that 
the problem with additional symmetry is not characteristic for the ADM
approach only but it also afflicts the configuration space of the (usual)
second order (Lagrangian) approach.

Our results are of course only valid for the two particular models. Of these,
the torus may be the most pathological case that exists. The Kucha\v{r}
description of less pathological models with additional symmetry may,
therefore, be regular. If not, and if the residual pathology is very
disturbing, one can still truncate the model. More cases ought to be looked
at, and some attempts at proofs of some general theorems ought to be done.
This is left for future papers.

\subsection*{Acknowledgments}
Discussions with A.~Fischer, K.~V.~Kucha\v{r}, V.~Moncrief and B.~G.~Schmidt
have been very helpful. P. H. thanks to Max-Planck-Institute for Gravitational
Physics in Golm for nice hospitality and support. The work has also been
supported by the Swiss Nationalfonds and by the Tomalla Foundation, Zurich.

\end{document}